%% file: life-sns_project_EPJC.tex
\documentclass[twocolumn,epjc3]{svjour3}
\RequirePackage[T1]{fontenc}

\smartqed  

\usepackage{cancel}
\usepackage{dcolumn}
\usepackage{esvect}
\usepackage{siunitx}

\usepackage{graphicx}%
\usepackage{multirow}%
\usepackage{amsmath,amssymb,amsfonts}%
\usepackage{mathrsfs}%
\usepackage[title]{appendix}%
\usepackage{xcolor}%
\usepackage{textcomp}%
\usepackage{manyfoot}%
\usepackage{booktabs}%
\usepackage{algorithm}%
\usepackage{algorithmicx}%
\usepackage{algpseudocode}%
\usepackage{listings}%

\RequirePackage{graphicx}
\usepackage{newtxtext,newtxmath} 
\RequirePackage{flushend}
\RequirePackage[numbers,sort&compress]{natbib}
\RequirePackage[colorlinks,citecolor=blue,urlcolor=blue,linkcolor=blue]{hyperref}

\newcommand{\Mo}[1]{$^{#1}$Mo}

\newcommand{\um}[1]{\SI{#1}{\micro \meter}}

\graphicspath{ {figs/} }

\begin{document}

\newcolumntype{d}[1]{D{.}{\cdot}{#1} }
\title{LiFE-SNS: LiF Experiment for keV-scale Sterile Neutrino Search}


\author{Y.C. Lee \thanksref{addr1,addr2} \and
        J.S. Chung\thanksref{addr1}         \and
        S.H. Choi\thanksref{addr2}          \and
        J.A. Jeon\thanksref{addr1}          \and
        D.H. Hwang\thanksref{addr1}     \and
        C.S. Kang\thanksref{addr1}     \and
        H.B. Kim\thanksref{addr1}           \and
        Ho Jong Kim\thanksref{addr1}         \and
        Hyeok Jun Kim\thanksref{addr1}        \and
        H.L. Kim\thanksref{addr1}           \and
        M.B. Kim\thanksref{addr1}           \and
        S.C. Kim\thanksref{addr1}           \and    
        S.K. Kim\thanksref{addr2,e1}        \and
        W.T. Kim\thanksref{addr1}           \and  
        Y.H. Kim\thanksref{addr1,addr3,e2}  \and    
        Y.M. Kim\thanksref{addr1}     \and
        D.H. Kwon\thanksref{addr1,addr3}    \and    
        D.Y. Lee\thanksref{addr1}     \and
        H.J. Lee\thanksref{addr1}     \and
        S.H. Lee\thanksref{addr1}     \and
        S.W. Lee\thanksref{addr1}     \and
        H.S.~Lim\thanksref{addr1}           \and
        H.S.~Park\thanksref{addr4}          \and
        K.R.~Woo\thanksref{addr1,addr3}     \and
        J.Y. Yang\thanksref{addr1}          \and
        Y.S. Yoon\thanksref{addr4}          
}

\thankstext{e1}{e-mail: skkim@snu.ac.kr}
\thankstext{e2}{e-mail: yhk@ibs.re.kr}

\institute{Center for Underground Physics, Institute for Basic Science (IBS), Daejeon 34126, Republic of Korea\label{addr1}
          \and
          Department of Physics and Astronomy, Seoul National University, Seoul 08826, Republic of Korea\label{addr2}
          \and 
          IBS School, University of Science and Technology (UST), Daejeon 34113, Republic of Korea\label{addr3}        
          \and
          Korea Research Institute of Standards and Science, Daejeon 34113, Republic of Korea\label{addr4}
}

\date{Received: date / Accepted: date}

\maketitle

\begin{abstract}
The LiF Experiment for keV-scale Sterile Neutrino Search (LiFE-SNS) aims to probe sterile neutrinos through precision measurements of the tritium $\beta$ spectrum. Tritium nuclei are produced and embedded in LiF crystals via the ${}^{6}\mathrm{Li}(n,\alpha){}^{3}\mathrm{H}$ reaction, allowing thermal calorimetric detection of $\beta$ decays with magnetic microcalorimeters (MMCs) operated at millikelvin temperatures. 
We present the detector configuration, background studies, and calibration method, including modeling of position-dependent response and characterization of detector nonlinearity. We also discuss potential sources of systematic uncertainty relevant to the sterile-neutrino search. While the first phase of LiFE-SNS has been completed, this paper focuses on calibration and detector characterization. The achieved performance enables precision $\beta$-spectrum measurements, and projected sensitivities indicate competitive reach in the keV mass region.
\end{abstract}

\input{sec1_intro.tex}

\input{sec2_beta.tex}

\input{sec3_experiments.tex}
\input{sec4-1_results.tex}

\input{sec4-2_discussions}

\input{sec5_systematics.tex}

\input{sec6_conclusion.tex}

\input{appendix}

\section*{Acknowledgement}
This research is supported by Grant no. IBS-R016-A2.


\end{document}

%% file: sec1_intro.tex
\section{Introduction} 

Sterile neutrinos are hypothetical particles motivated by the neutrino Minimal Standard Model ($\nu$MSM), an extension of the Standard Model that introduces three right-handed (sterile) neutrinos which mix only feebly with the active neutrinos~\cite{ASAKA2005151}. Through the seesaw mechanism~\cite{minkowski1977plb,gellmann2010book,yanagida1980ptp,mohapatra1980prl}, 
the heavy sterile states account for the small masses of the active neutrinos, whereas the mass of the lightest sterile state (the subject of this report) remains largely unconstrained.

Sterile neutrinos with keV-scale masses may realize warm dark matter (WDM)~\cite{bode2001aj}. Lower bounds from the Tremaine\hspace{0pt}–Gunn argument require masses above $\sim$1\,keV while various production scenarios for WDM and astrophysical experiments favor a mass range of 
$4\!-\!50$\,keV~\cite{adhikari2017jcap,yeche2017jcap,degouva2020prl,boyarsky2008mnras,laine2008jcap}.
Reports of a 3.5\,keV spectral feature, potentially from $\sim$7\,keV sterile neutrinos~\cite{bulbul2014apj,boyarsky2014prl,cappelluti2018apj,roach2020prd}, remain under debate~\cite{sicilian2022apj,dessert2024apj,foster2021prl,krivonos2024prl}, and forthcoming XRISM observations are expected to clarify this issue~\cite{tashiro2025asj,xrism2025apjl}.

Non-observation of such X-ray signatures constrains the active–sterile mixing angle to roughly $\sin^{2}2\theta \sim 10^{-12}$ for 10\,keV sterile neutrinos accounting for the full dark matter abundance.
However, these limits are strongly model-dependent due to uncertainties in early universe production mechanisms and in the radiative decay rate itself~\cite{gelmini2004prl,yaguna2007jhep,abazajian2023prd,benso2019prd}. A more direct and less model-dependent approach is $\beta$-decay spectroscopy, where emission of a sterile neutrino with a mass below the decay $Q$ value leads to a characteristic distortion in the $\beta$ spectrum.

The LiF Experiment for keV Sterile Neutrino Search (LiFE-SNS) aims to exploit this approach using tritium $\beta$ decay measured at mK temperatures. LiF crystals containing 
$^3$H produced via the Li(n,$\alpha$)$^3$H reaction serve as absorbers, and magnetic microcalorimeters (MMCs) record the deposited energy with high resolution~\cite{enss2000jltp,enssbook:mmc,kempf2018jltp,yhkim2022sust}. 
This report outlines the detection principle, technical performance, and current status of the LiFE-SNS project.

\begin{figure} [t]
\centering
\includegraphics[width=0.8\columnwidth]{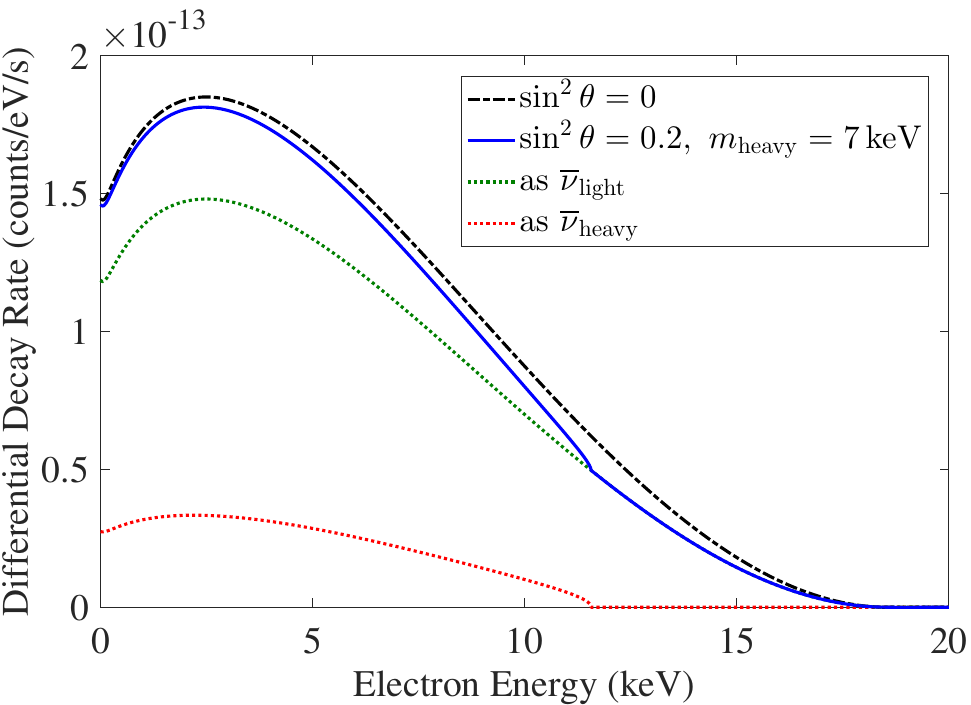}
\caption{
 Expected spectrum of $^3$H $\beta$ decay. The black dotted line represents the $\beta$ spectrum with no mixing with the heavy neutrino state. The blue solid line shows the expected decay spectrum of the $\beta^-$ electron from a $^3$H as the sum of the light (green dotted line) and heavy (red dotted line) terms. The rates are calculated with an assumption of $\sin ^2 \theta = 0.2$ and $m_\mathrm{heavy} = 7$\,keV. 
} 
\label{fig:non_rel_tri_spec}
\end{figure}

%% file: sec2_beta.tex
\section{Sterile neutrino search from $\beta$-spectrum measurement}

Beta decay proceeds via the weak interaction, producing an electron and an electron antineutrino that share the total decay energy $Q$. The electron energy ranges from zero to an endpoint shifted by the neutrino mass from the $Q$ value. In the three–active-neutrino framework, the effective electron-neutrino mass is
\begin{equation}
m_{\mathrm{light}}^{2}=\sum_{i=1}^{3} |U_{ei}|^{2} m_{i}^{2},
\end{equation}
with $U$ the PMNS matrix. Because the masses and mass differences are nearly zero, the three light mass states can be treated effectively as a single state. 
If a heavier sterile state mixes with the active one,
\begin{equation}
\bar{\nu}_{e}=\cos\theta\,\bar{\nu}_{\mathrm{light}}+\sin\theta\,\bar{\nu}_{\mathrm{heavy}},
\end{equation}
beta decay can also emit a heavy neutrino. The observed spectrum becomes
\begin{equation}
\frac{d\Gamma_{\rm tot}}{dE}
=\cos^{2}\theta\,\frac{d\Gamma}{dE}(m_{\mathrm{light}})
+\sin^{2}\theta\,\frac{d\Gamma}{dE}(m_{\mathrm{heavy}}),
\end{equation}
so a sterile state appears as a kink at $E = Q - m_{\mathrm{heavy}}$. Its position and amplitude reveal the sterile-neutrino mass and mixing angle. The spectral shape $d\Gamma/dE$ is further discussed in section \ref{sec:systematics}.

\begin{figure} [t]
\centering
\includegraphics[width=1.0\columnwidth]{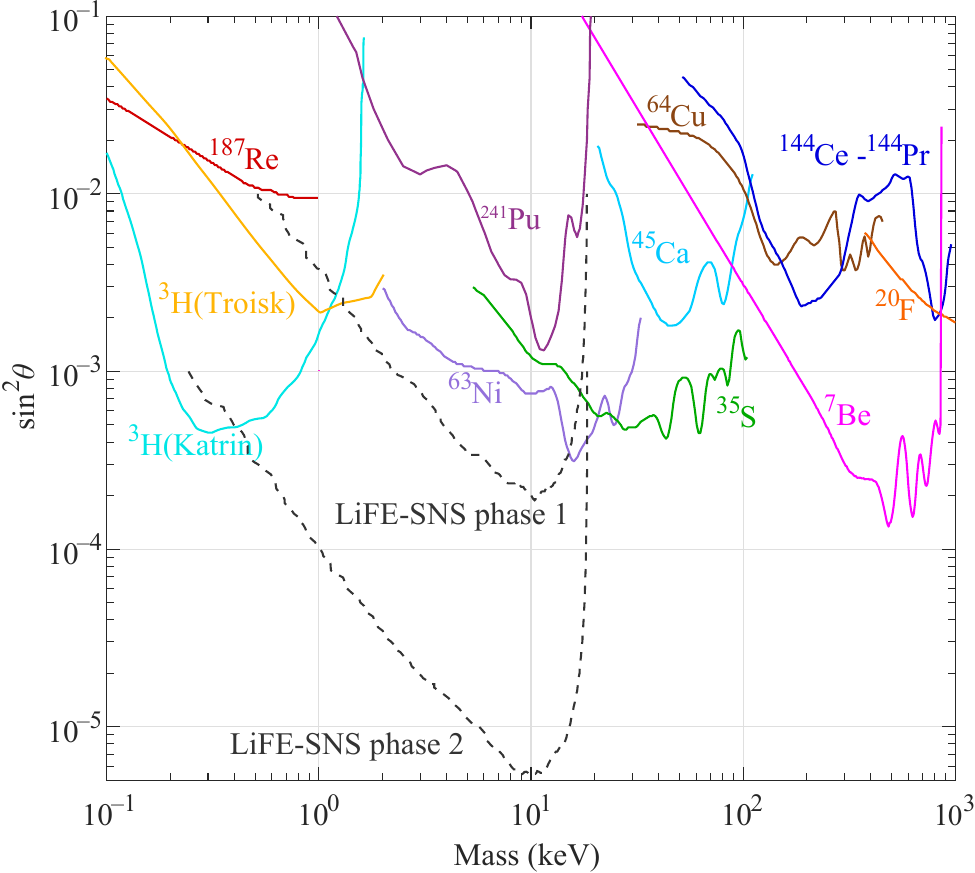}
\caption{
Upper bounds of the experimental limits for sterile neutrino searches in keV scale from $\beta$ decay measurements. Their references are listed in Table~\ref{tab:beta_isotopes}. The dotted lines are the expected sensitivities of the LiFE-SNS phases. Phase~1 corresponds to $6\times10^{8}$ $\beta$ events from a two-channel four-month exposure, while Phase~2 assumes $10^{12}$ $\beta$ events with increased channel count over a multi-year period. 
} 
\label{fig:bounds}
\end{figure}

Figure~\ref{fig:non_rel_tri_spec} illustrates this feature for tritium $\beta$ decay using an unrealistically large mixing with $m_{\mathrm{heavy}} = 7$\,keV to make the spectral kink visible.
Numerous experiments employing different $\beta$ emitters have searched for such spectral distortions, as listed in Table~\ref{tab:beta_isotopes}. Their exclusion limits, summarized in Fig.~\ref{fig:bounds}, constrain the mixing angle to the $10^{-4}$--$10^{-3}$ range over 0.1--1000\,keV. 
To date, the most sensitive limits were found from BeEST using $^{7}$Be electron capture (100--900\,keV)~\cite{friedrich2021prl}, a magnetic $\beta$ spectrometer using $^{63}$Ni and $^{35}$S sources (1--100\,keV)~\cite{holzschuh1999plb,holzschuh2000plb},  and KATRIN using $^{3}$H for sub-keV masses~\cite{aker2023epjc}. 
Notably, the 1--10\,keV region relevant for LiFE-SNS remains comparatively weakly constrained.

\begin{table}[b]
\caption{Beta decaying isotopes used for sterile neutrino searches with Q values less than 1\,MeV. }
\label{tab:beta_isotopes}
\begin{tabular}{ l l l } 
\toprule 
Isotope  & Q (keV) & Experiment \\ 
\midrule 
$^3$H     & 18.591  & \cite{Abdurashitov2017,aker2023epjc}, this work       \\ 
$^{7}$Be\footnotemark[1]   & 861.89  & \cite{friedrich2021prl}  \\ 
$^{35}$S  & 167.33  & \cite{holzschuh2000plb}     \\ 
$^{45}$Ca & 235.8   & \cite{derbin1997jetp}       \\ 
$^{63}$Ni & 66.945  & \cite{holzschuh1999plb,dylee2025ieeetas}  \\ 
$^{131}$Cs\footnotemark[1]  & 355  & \cite{martoff2021hunter}  \\ 
$^{64}$Cu & 579.6   & \cite{schreckenbach1983plb}  \\ 
$^{144}$Ce--$^{144}$Pr & 318.7--2997.5  & \cite{derbin2018jeptl} \\ 
$^{187}$Re & 2.469  & \cite{galeazzi2001prl}      \\ 
$^{241}$Pu & 20.78  & \cite{clee2025arxiv}     \\ 
\bottomrule 
\end{tabular}
\\
\footnote{1}{Electron capture decay.}
\end{table}

%% file: sec3_experiments.tex
\section{Experiments}

\subsection{Tritium generation} 
\label{subsec:tritiumGen}

$^3$H can be generated and embedded in a LiF crystal from an $^6\mathrm{Li}(n,\alpha)^3\mathrm{H}$ reaction 
\begin{equation}
{^{6}\mathrm{Li}} + n \rightarrow  {^{4}\mathrm{He}}  + {^{3}\mathrm{H}}  
\end{equation}
with a $Q$ value of 4.78\,MeV. The pair of triton and alpha particle share this energy by 2.73\,MeV and 2.05\,MeV, respectively. 
The natural abundance of $^6$Li is 7.6\% and  $^6$Li has a large cross-section for thermal neutrons. The mean free path of thermal neutrons is about 2.3\,mm in a LiF crystal. 

The neutron activation of LiF crystals was conducted at the laboratory storage facility of the Korea Research Institute of Standards and Science (KRISS)~\cite{sckang2023jkps}. This facility houses three neutron sources: two AmBe sources and one $^{252}$Cf source encased in polyethylene bricks. The expected neutron emission rate from these sources ranges from 1 to 2 $\times$10$^{7}$~n/s. A small sample loading area is positioned approximately 17\,cm from the neutron sources, ensuring that neutrons are sufficiently thermalized within the polyethylene before reaching the LiF crystals.

\begin{figure} [t]
\centering
\includegraphics[width=1.0\columnwidth]{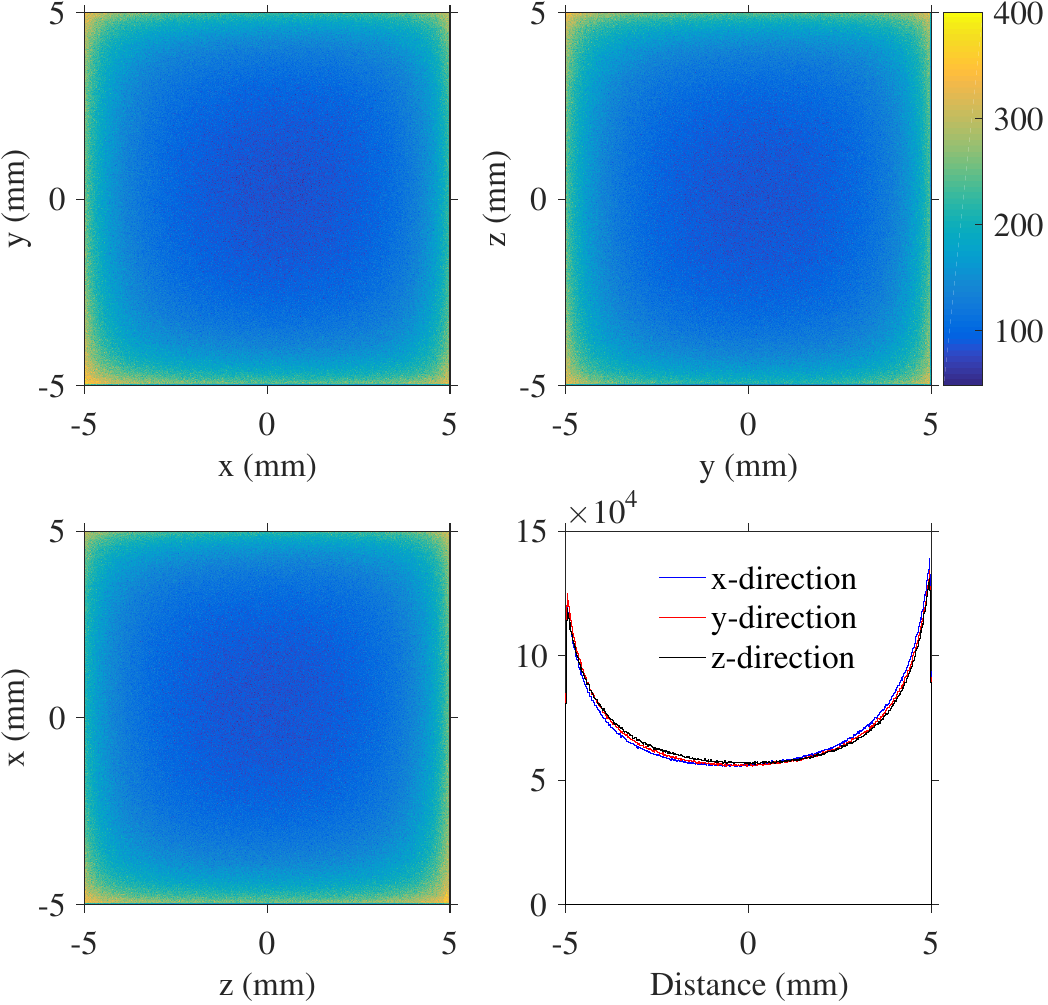}
\caption{Expected $^3$H distribution in a 1$\times$1$\times$1\,cm$^3$ LiF crystal exposed to a thermal neutron flux at the sample loading area of the KRISS neutron source storage. The 3D distribution is projected to the \textit{xy}, \textit{yx} and \textit{zx} planes, as well as along the \textit{x}, \textit{y}, and \textit{z} directions, with a bin size of 20\,\um{}. 
} 
\label{fig:tritium-dist}
\end{figure}

A Geant4 simulation was performed to estimate the $^3$H spatial distribution within the crystals. The simulation modeled neutron generation, scattering, and the subsequent capture processes, designed geometries for the polyethylene bricks, neutron sources, and sample placement. Approximately 35 million $^3$H nuclei were generated in the crystal from 1.3$\times$10$^{11}$ initial neutrons. This corresponds to an estimated $^3$H $\beta$ activity of about 20\,Bq for a 1\,cm$^3$ LiF cube exposed for one week. As shown in Fig~\ref{fig:tritium-dist}, the resulting $^3$H distribution is relatively depleted in the central region compared to the edges and surfaces, while it exhibits minimal directional dependence.

The simulation also reveals that the $^3$H nuclei are displaced by approximately 33\,\um{} from the original $^6$Li lattice sites following neutron capture. In cases where capture occurs near the surface, some $^3$H nuclei escape the crystal before fully dissipating their kinetic energy. The depth profile indicates that the $^3$H density peaks around 35\,\um{} beneath the surface.

On the other hand, electron spin resonance studies of neutron-activated LiF crystals from the 1960s and 1970s suggest that these nuclei tend to occupy interstitial lattice sites 
and remain stable unless annealed above 200\,$^\circ$C~\cite{kaplan1963pr,kazumata1973esr,kamikawa1980pss}. 
The temperature-dependent diffusion constant was found to be approximately \SI{0.03}{\micro  \meter^2 \per yr} at 293\,$^\circ$C~\cite{cohen1965diffusion,nullmeyer2017apl}.
 The locations of $^3$H in LiF are expected to change negligibly during the few months of sample preparation at room temperature after neutron activation and during the few years of the data collection period at mK temperatures.

\subsection{Sensor technology}

A low temperature detector (LTD) is typically made up of an absorber and a temperature sensor (in this work, a LiF crystal and an MMC sensor) that are weakly thermally coupled to a heat bath at a temperature of $\mathcal{O}(10\,\mathrm{mK})$. When the thermal connection between the absorber and sensor is much stronger than that to the heat both, the maximum temperature rise $\delta T_\mathrm{max}$ resulting from an energy input $E$ can be estimated as
\begin{equation}
\delta T_\mathrm{max} = \frac{E}{C_\mathrm{tot}}
\end{equation}
where $C_\mathrm{tot}$ is the sum of the heat capacities of the detector components, including the absorber and sensor. This expression follows from a simplified thermal model under the condition $\delta T_\mathrm{max}/T_0 \ll 1$, where $T_0$ is the bath temperature.

MMC sensors, one of the most established technologies for high-resolution cryogenic particle detection~\cite{enss2000jltp,enssbook:mmc,kempf2018jltp}, employ a paramagnetic material, typically silver or gold doped with a small erbium concentration, as the temperature sensor. Magnetic signals are read out via a superconducting circuit using a dc-SQUID, which converts changes in magnetic flux $\delta\Phi$ originating from a temperature increase into a voltage signal. The flux change can be written as
\begin{equation}
    \delta \Phi_\mathrm{max} = \xi  \frac{\partial M}{\partial T}  \delta T_\mathrm{max} = \xi  \frac{\partial M}{\partial T}  \frac{E}{C_\mathrm{tot}}
\end{equation}
where $\partial M/\partial T$ indicates the temperature dependence of the sensor magnetization and $\xi$ is a proportionality parameter that includes the sensor geometry with respect to the superconducting pickup coil and the inductance parameters of the circuit.   

MMC sensors can be microfabricated using established micropatterning techniques for superconducting circuits and paramagnetic alloys~\cite{burck2008jltp,fleischmann2009aip,wsyoon2014jltp}. The sensors used in this work were developed for the AMoRE project~\cite{sgkim2021ieeetas}, which investigates neutrinoless double-beta decay of \Mo{100}~\cite{amore_tdr,amore2019epjc,amore2025prl}.

During detector operation, a persistent current is injected into a superconducting Nb coil on the MMC chip to provide a stable magnetic field to the sensor material. The signal then appears in the SQUID output without heat dissipation and without galvanic contact to the sensor. 

\begin{figure} [t]
\centering
\includegraphics[width=1.0\columnwidth]{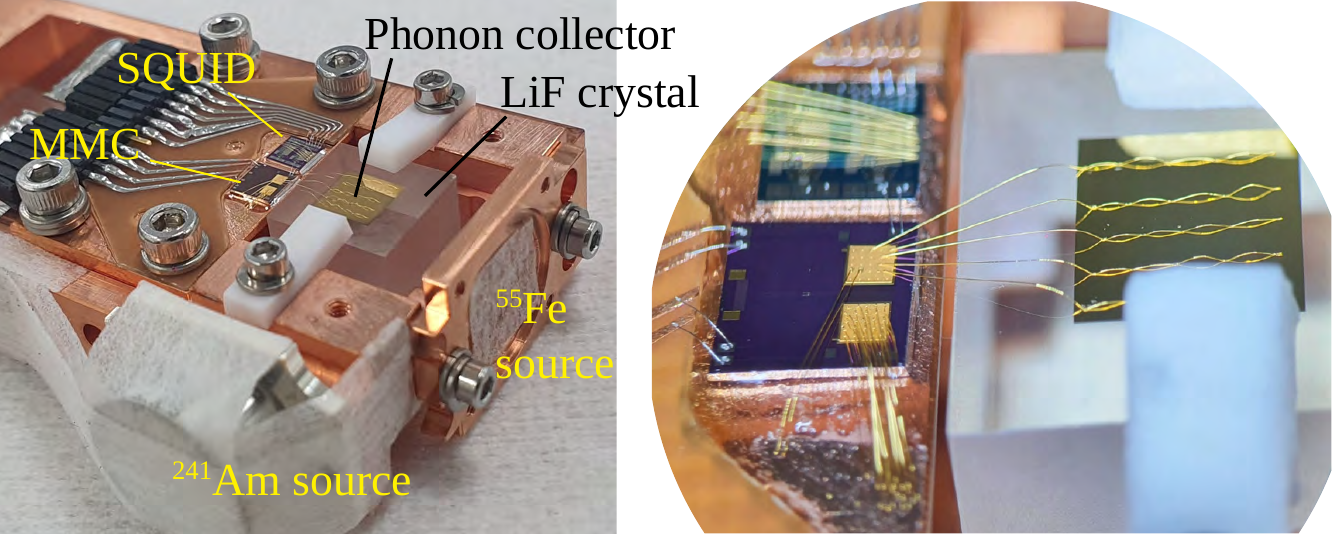}
\caption{
A LiFE-SNS detector setup composed of a LiF crystal and an MMC sensor. An $^{241}$Am source with a silver collimator and an $^{55}$Fe source were introduced for energy and position calibration. This picture shows the source configurations of Exp.~5. 
} 
\label{fig:setup}
\end{figure}

\subsection{Detector setup}

Figure~\ref{fig:setup} shows photographs of a detector setup used for the LiFE-SNS project with a LiF crystal and an MMC sensor. A similar configuration was used in the first proof-of-principle measurement for LiFE-SNS~\cite{yclee2022jltp}. Originally, this detector design was developed to investigate the phonon and scintillation properties of various crystals at mK temperatures~\cite{hlkim2018tns,hlkim2020jltp,hlkim2020nima}.

The crystal placed in the copper frame was fixed with two Teflon clamps. Four small pieces of Teflon sheets were placed between the cubic crystal and the copper holder at the four corners of the bottom surface, not shown in Fig.~\ref{fig:setup}. 
The crystal has a 5\,mm$\times$5\,mm$\times$300\,nm gold film deposited on the top surface. 
The gold film is an intermediate thermal component that serves as a phonon collector to make a strong thermal connection to an MMC sensor with gold bonding wires~\cite{yhkim2004nima,hlkim2023epjp}.
Five 25\,\um{} gold wires are bonded on the film and the MMC sensor to make their thermal connection. Each bonding wire is extended from one side to the other with several bonding points on the film, as shown in the zoom-in picture. This extension enhances the thermal conductance in the horizontal direction across the film.   
 The MMC sensors used in the present experiment were fabricated in a Ag:Er(400\,ppm) batch. 

The detector setup accommodates two calibration sources, $^{55}$Fe (emitting 5.9\,keV and 6.5\,keV X-rays) and $^{241}$Am (emitting several lines of X- and $\gamma$-rays including 59.54\,keV $\gamma$-rays), which can be positioned on all sides of the crystal. The X- and $\gamma$-rays from the $^{241}$Am source were guided by a silver (or copper) collimator, which produces additional characteristic X-ray lines in the 20\,keV (or 8\,keV) region.

\begin{figure} [t]
\centering
\includegraphics[width=0.9\columnwidth]{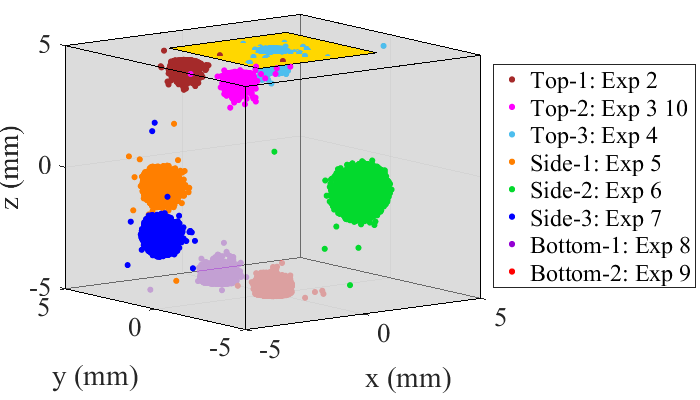}
\caption{
Simulated positions of 5.9 keV X-ray events (Mn K$_\alpha$) absorbed in the LiF crystal from various $^{55}$Fe source locations used in Exp.~2--10. Geant4 simulations were performed using the designed detector geometry, including the source, collimator, and crystal, to model the X-ray absorption in the crystal.  
} 
\label{fig:6kev-pos}
\end{figure}


\begin{table}[b]
\caption{Eleven experimental runs with various calibration sources}
\begin{tabular}{llll}
\toprule
Exp.                              & Refrigerator & Calibration sources       & Section        \\ 
\midrule
1                                     & DR           & $^{55}$Fe, $^{241}$Am with Ag collimator & \ref{subsec:basic-performance}, \ref{subsec:gamma-background} \\ 
\midrule
2--9 & ADR          & $^{55}$Fe, $^{241}$Am with Ag collimator & \ref{subsec:position-calib}, \ref{appendix} \\
\midrule
10                                    & ADR          & $^{55}$Fe, $^{241}$Am with Cu collimator & \ref{subsec:position-calib}, \ref{appendix} \\ 
\midrule
11\footnotemark[2]     & DR           & $^{55}$Fe, (Ag, W)\footnotemark[3] & \ref{exp11} \\
\bottomrule 
\end{tabular}
\label{tab:exp}
\\
\footnote{2}{After neutron activation.} \\
\footnote{3}{Characteristic X-rays of Ag and W could be activated from external $\gamma$ sources during the calibration run.}
\end{table}

\subsection{Experiment Runs}

This paper reports a series of 11 experimental runs conducted using a detector setup in two different refrigerators: an adiabatic demagnetization refrigerator (ADR) and a dilution refrigerator (DR). Table~\ref{tab:exp} summarizes the refrigerators and calibration sources used in each experiment.

The ADR was employed for runs that required frequent thermal cycling. Eight runs (Exp. 2–9) were carried out with a $^{55}$Fe source collimated at various positions on different surfaces of the LiF crystal relative to the phonon collector film. The $^{241}$Am source remained fixed at the side of the crystal during these runs, as shown in Fig.~\ref{fig:setup}. 

\begin{figure} [t]
\centering
\includegraphics[width=0.85\columnwidth]{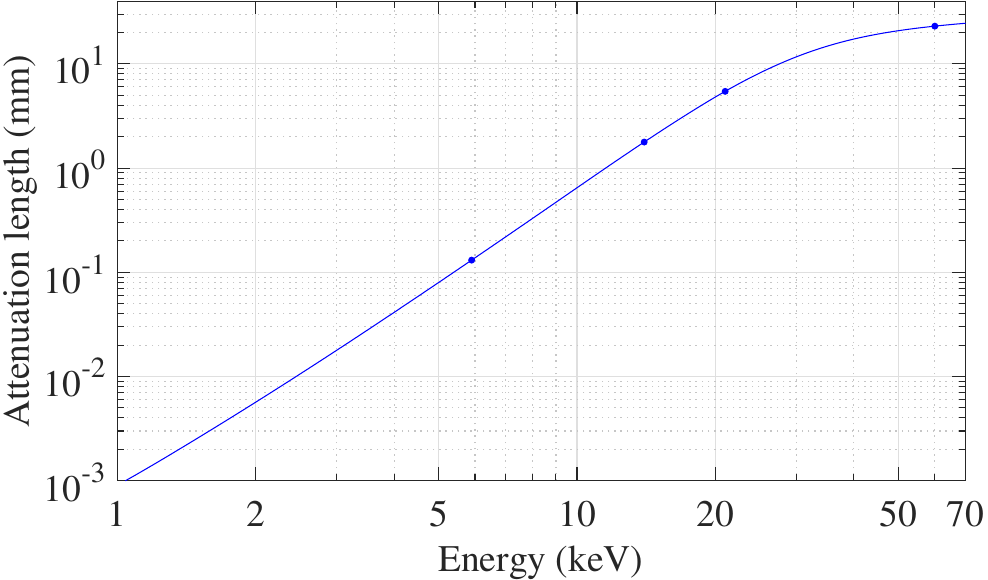}
\caption{
Photon attenuation length in a LiF crystal. Photoelectric absorption and Compton scattering are included in the total attenuation calculation. The corresponding attenuation lengths for major X-ray lines at 5.9, 14,  21, and 60\,keV are approximately 0.13, 1.8, 5.3, and 23\,mm, respectively.
} 
\label{fig:att-length}
\end{figure}

In Exp. 10, the $^{241}$Am source was relocated to the bottom of the crystal, while the $^{55}$Fe source remained at the same position as in Exp. 3--near the phonon collector film on the top surface. Figure~\ref{fig:6kev-pos} shows the expected event distributions for various $^{55}$Fe source locations with respect to the crystal and phonon collector film. Note that the photon attenuation lengths in LiF for the relevant energy range are shown in Fig.~\ref{fig:att-length}. 
The distribution of X-ray events in the crystal depends on both the photon attenuation at the corresponding energy and the source–collimation geometry.

Two additional measurements, Exp.~1 and Exp.~11, were conducted in a dry DR. 
This DR system featured a 10-cm-thick external Pb shield around the cryostat and an additional 10-cm-thick internal Pb shield between the mixing chamber plate and the detectors. 
Exp.~1 focused on characterizing the detector, including energy calibration and resolution performance. 
Another aim of this run was to evaluate the $\gamma$ background measured with a 1$\times$1$\times$1\,cm$^3$ LiF crystal in the environmental measurement condition.

After the characterization measurements, Exp.~1--10, two LiF crystals were neutron activated at the KRISS site for two weeks. A four-month measurement campaign (Exp.~11) was carried out in the DR using two detector setups as part of the first phase of the LiFE-SNS experiment. This report includes results from a 13-day calibration campaign using the same crystal as in Exp.~1–10.

%% file: sec4-1_results.tex
\section{Results and Discussions}   

 

\subsection{Amplitude Determination}

The detector signals were digitized with 16-bit resolution ($\pm$5\,V) at a sampling rate of $5 \times 10^5$ samples/s and continuously stored for offline analysis. 
A bandpass filter with an optimized frequency window of [100 1500]\,Hz passband was applied~\cite{iwkim2018jltp}. Each trigger captured a 2\,ms-long window, including a 0.36\,ms pre-trigger region.

For detector characterization, a noise-free signal template was constructed averaging full-absorption events of 59.54-keV $\gamma$-rays from the $^{241}$Am source during experimental runs (Exp. 1--10), which were dedicated to investigating the initial detector performance and position dependence. Specifically, the signal amplitude was determined using a least-squares fit to the scaled template in the time domain~\cite{hslim2024jltp}.

\begin{figure} [t]
\centering
\includegraphics[width=0.8\columnwidth]{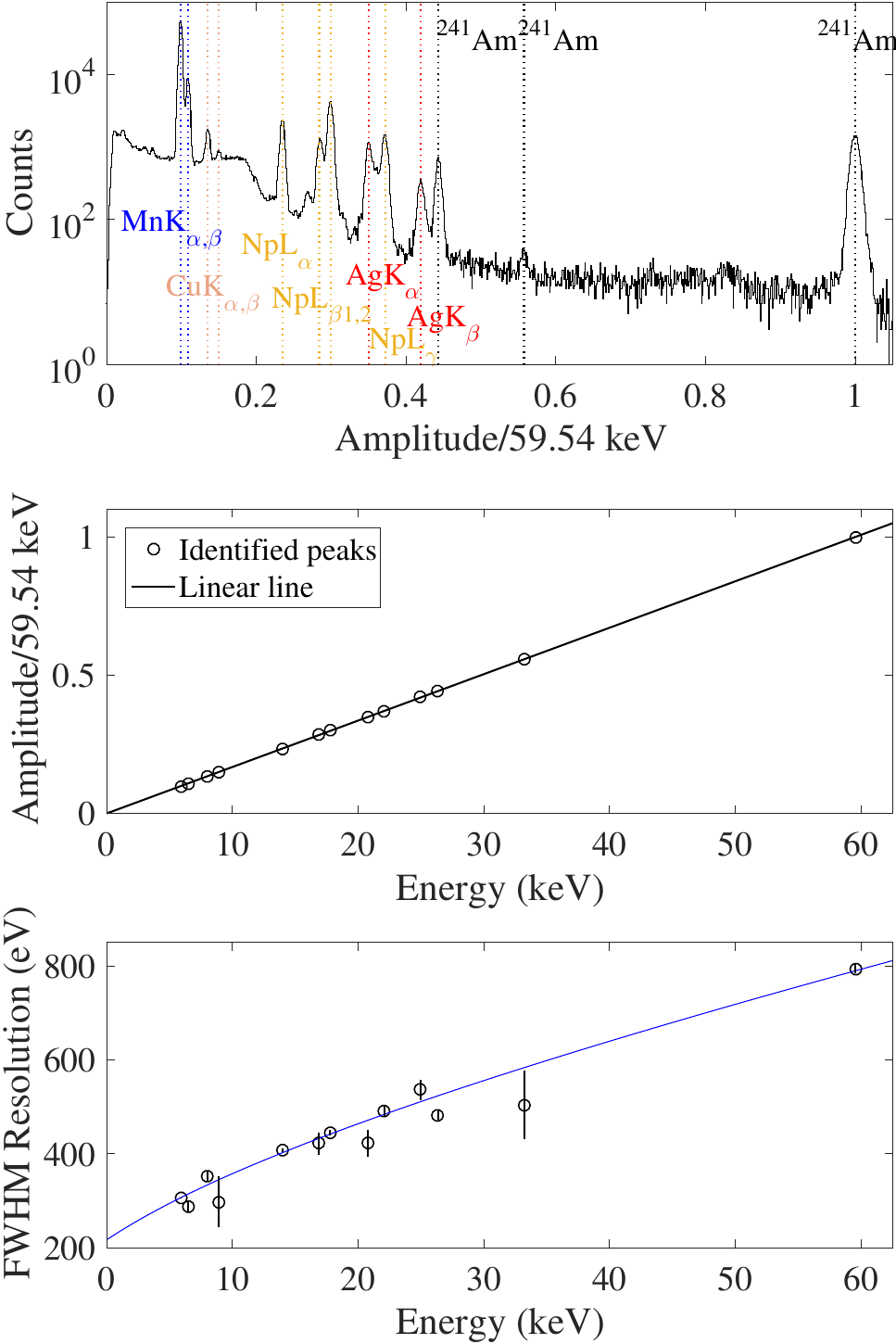}
\caption{
({\textit{Top}}) A calibration spectrum of a LiF detector in Exp.~1. The origins of the peaks are specified with the vertical lines. ({\textit{Middle}}) The amplitude alignments to a straight line with respect to 59.54\,keV events. 
({\textit{Bottom}}) Energy resolution of the peaks. 
The solid line guides a quadratic fit of the resolution squared ($\sigma ^2$) to the energy.
These plots show the basic detector performance describing the energy calibration and resolution based on a simplified model. Further details should be taken into account for detector nonlinearity and event locations of photon absorption sites in the LiF crystal.
}
\label{fig:linear}
\end{figure}

\subsection{Basic performance}
\label{subsec:basic-performance}

The top plot in Fig.~\ref{fig:linear} shows a histogram of the measured amplitudes 
from Exp.~1. As shown in the middle plot, peak identification was straightforward due to the high linearity of the MMC sensor response~\cite{gbkim2017app, iwkim2017sust, hlkim2023epjp}. All observed X-ray peaks could be traced to either the calibration sources or X-ray fluorescence from nearby materials of silver and copper. The peaks of the characteristic lines in the amplitude scale closely followed a linear trend with respect to the known energy values. 

Although the pulse amplitudes generally follow a linear response, nonlinear and position\hspace{0pt}-dependent deviations were observed. These effects are discussed in Section~\ref{subsec:position-calib}, where we address the need for accurate energy and position calibration.

The bottom plot in Fig.~\ref{fig:linear} presents the energy resolution of the detector system, which exhibits an energy-dependent trend. Full width at half maximum (FWHM) resolutions of 200--500\,eV were obtained in the energy region of interest for the keV-scale sterile neutrino search using $^3$H $\beta$ decay measurements.

\subsection{Environmental gamma background}
\label{subsec:gamma-background}

\begin{figure} [t]
\centering
\includegraphics[width=1.0\columnwidth]{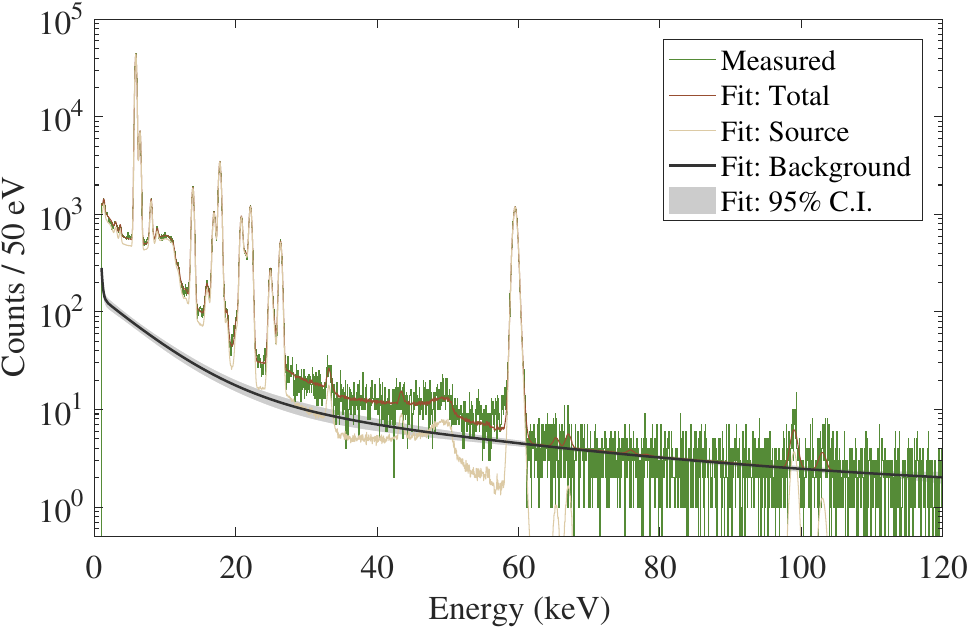}
\caption{%
Measured and fitted spectra from the detector characterization in Exp.~1, corresponding to a live-time of approximately 14 days. The total fit represents the sum of energy spectra from an environmental $\gamma$ background and the two internal calibration sources, $^{241}$Am and $^{55}$Fe. The Geant4 simulation incorporated the materials and geometry of the detector system to model X-ray and photoelectron emissions accurately. 
} %
\label{fig:gamma-bkg}
\end{figure}

Although the spectrum shown in Fig.~\ref{fig:linear} primarily originates from the detector response to the internal sources, the environmental $\gamma$ background also contributes to the measured events. A Geant4 simulation was conducted to estimate the energy deposited in the LiF crystal by the sources. The simulation incorporated a detailed implementation of the detector geometry and materials, including the collimators and sources.

Figure~\ref{fig:gamma-bkg} shows the measured spectrum overlaid with a combined fit consisting of the simulation result and a background function composed of three exponential decay components and a constant term. The fit agrees well with the measured data across the energy range of 1--120\,keV, to ensure sufficient statistics in the region unaffected by the sources. A simple quadratic function without a constant term was used for energy calibration. The resulting background spectrum was adopted for the $^3$H $\beta$ spectrum analysis, as both Exp.~1 and Exp.~11 were conducted under identical environmental conditions in the DR system. Position calibration, discussed in the following subsection, was not included in this background estimation. The error from the calibration is expected to be significantly smaller than the statistical uncertainty of the measured spectrum.

\begin{figure} [t]
\centering
\includegraphics[width=1.0\columnwidth]{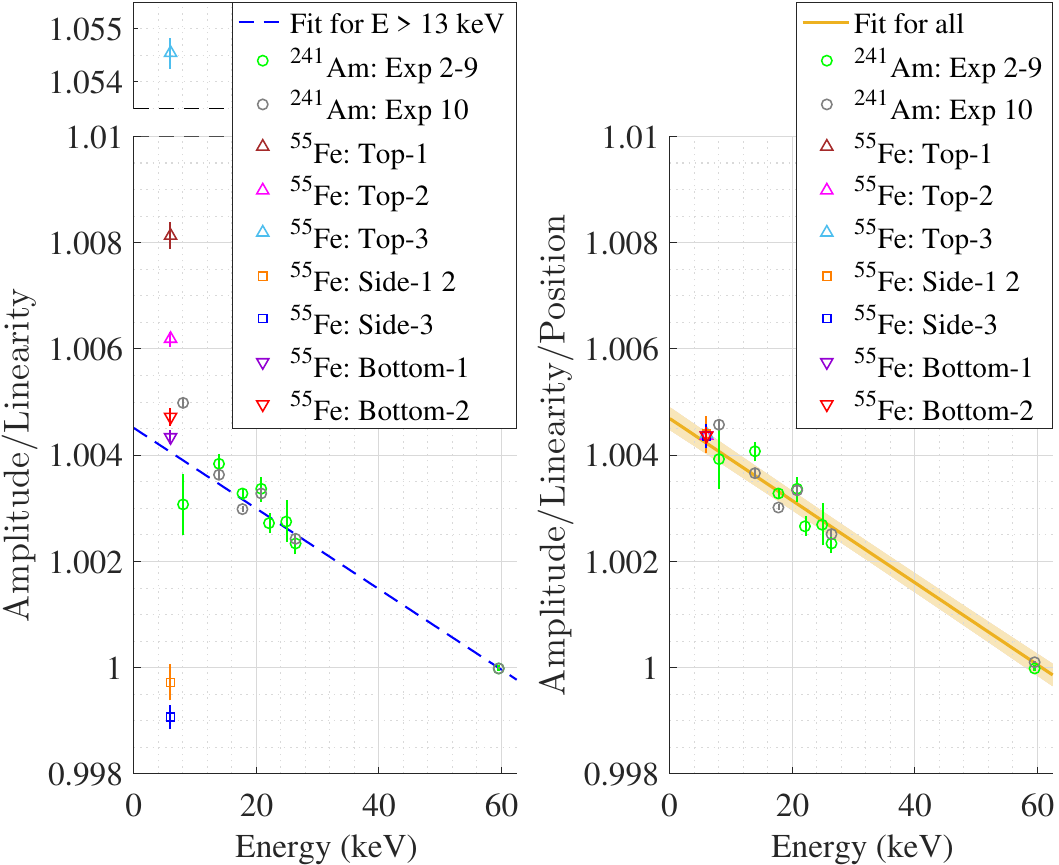}
\caption{%
Before (\textit{left}) and after (\textit{right}) applying position calibration to the signal amplitudes for various energies and source locations. 
Position-dependent effects are more pronounced at low energies due to the localized distribution of events relative to the gold film, resulting from photon attenuation of the collimated X-rays.
The dashed line shows the quadratic fit to peak amplitudes above 13\,keV, representing the intrinsic energy-dependent nonlinearity. 
After applying the position correction, all data points align with the solid line, which represents the updated universal quadratic calibration function. The shaded region denotes the 95\% confidence interval of the fit.
} %
\label{fig:pe-cal}
\end{figure}

\subsection{Position calibration}
\label{subsec:position-calib}

In general detector characterization, nonlinear behavior can be incorporated into the energy calibration function. MMC-based calorimeters typically require only a minor correction to account for deviations from perfect linearity. In the present analysis, performed prior to neutron activation to characterize the detector performance, we adopt a quadratic function without a constant term to describe the energy dependent response. Such deviations become more apparent when plotting signal amplitudes normalized by the ideal linear response~\cite{gbkim2017app,hlkim2020jltp}.

In addition to this energy-dependent nonlinearity, the detector exhibits position\hspace{0pt}-dependent behavior with respect to the phonon collector film. The left panel of Fig.~\ref{fig:pe-cal} shows the measured amplitudes of the X-ray peaks observed in Exp.~2--10, which were conducted with various source positions. The vertical axis shows the signal amplitude normalized to a one-point calibration using the 59.54\,keV events.

The blue dashed line represents the best-fit quadratic function to data points above 13\,keV, reflecting a small but measurable nonlinearity.
However, the amplitudes for the $^{55}$Fe events deviate from this trend.  
For instance, amplitudes are larger for source positions above the crystal (Exp.~2--4) compared to those below (Exp.~8 and 9), and the amplitudes from below exceed those from the side positions (Exp.~5--7).
The two side runs (Exp.~5 and 7) resulted in no significant difference in their amplitudes, corresponding to a single data point in the left plot. Notably, events from Top-3 (Exp.~4), where X-rays are absorbed just beneath the phonon collector film, exhibit significantly larger amplitudes than all other cases, which otherwise fall within a 1\% deviation. Some events absorbed directly in the gold film are distinguishable by their signal rise time and are excluded from this analysis.

Generally, pulse amplitudes can be modeled as a function of both energy $E$ and event position $(x,y,z)$ in the crystal. Assuming separability of the position and energy dependencies, the amplitude can be expressed as:
\begin{equation}
AMP  = F(x,y,z)   \cdot \mathcal{E}(E) = F(x,y,z)   \cdot (a E^2 + b E),
\label{eq:amp}
\end{equation}
where $\mathcal{E}(E)$ is approximated by a quadratic form without a constant term, as this matches the observed behavior of high-energy events. The position-dependence function $F(x,y,z)$ and the energy calibration parameters $a$ and $b$ can be extracted from measurements of the $^{55}$Fe peaks and additional full-absorption peaks originating from the $^{241}$Am source, across different source locations in Exp.~2--10. A total of 21 groups of data were used to evaluate the position-dependent response, and the detailed analysis is provided in Appendix~\ref{appendix}.

The right plot in Fig.~\ref{fig:pe-cal} shows the corrected amplitudes for all calibration lines. Most data points now align with the fitted quadratic calibration curve, demonstrating successful suppression of the position\hspace{0pt}-dependent effects. Notably, the 5.9-keV events from Exp.~4 (i.e. Top-3), which previously exhibited significant deviations, are now brought into agreement with the unified calibration. 

A key result of this analysis is that, with sufficient knowledge of the event position distribution $(x,y,z)$, the mean position\hspace{0pt}-dependent correction factor $\overline{F(x,y,z)}$ (as described in Appendix \ref{appendix}) can be robustly determined and applied to a measured spectrum to improve energy calibration accuracy.

%% file: sec4-2_discussions.tex
\begin{figure} [t]
\centering
\includegraphics[width=.8\columnwidth]{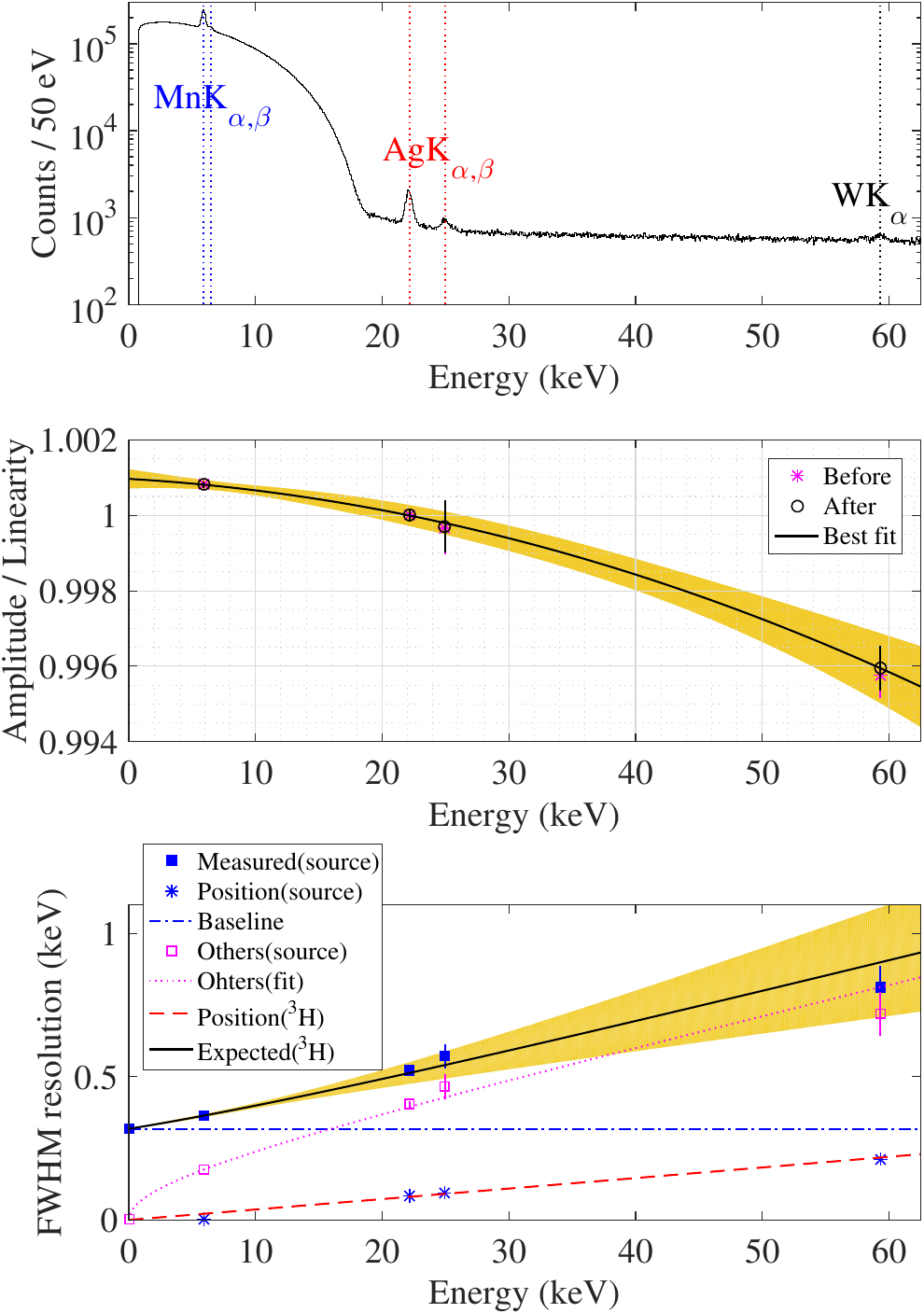}
\caption{ 
(\textit{Top}) Energy spectrum measured during the calibration campaign with the LiF detector setup.
(\textit{Middle}) Energy calibration plot before and after applying the position calibration. The vertical axis shows signal amplitudes divided by a linearized energy scale normalized to the expected amplitude for the 22.16\,keV Ag K$_\alpha$ template. The solid black line represents a third-order polynomial of the energy calibration function with 95\% confidence interval after applying the position correction.
(\textit{Bottom}) Energy resolution of the detector setup. The measured resolutions ($\sigma_\mathrm{measured}$) were obtained from the calibrated peaks. The black solid line shows the expected resolution with 95\% confidence interval for $^3$H $\beta$ decay events, incorporating the spatial distribution of $^3$H nuclei in the crystal (dashed line), the baseline resolution $\sigma_\mathrm{baseline}$ (horizontal line), and the other energy-dependent term $\sigma_\mathrm{others}$ (dotted line).
} 
\label{fig:e_cal_h3}
\end{figure}

\subsection{Calibration after neutron activation}
\label{exp11}

After neutron activation, two LiFE-SNS detector modules were assembled for a long-term measurement in the DR, designated as Exp.~11. A different calibration strategy was employed to measure the $^3$H $\beta$-decay spectrum across a wide energy range. For each setup, silver and tungsten plates were attached to the inside and outside surfaces of the aluminum sample-holder case. During the calibration campaign, several hundred thoriated-tungsten rods were placed around the cryostat, between the outer vacuum jacket and the external lead shield; these rods were removed prior to the actual $\beta$-spectrum acquisition. An $^{55}$Fe source was mounted at the \textit{Bottom-2} position of each LiF crystal as an internal calibration source, with significantly higher activity than that used in the previous experiments.

The dilution refrigerator system maintained the detector temperature at 24\,mK during the calibration campaign, followed by a four-month data-taking period (i.e. science run for the sterile-neutrino search). The temperature was stabilized by a two-stage control system~\cite{krwoo2022jltp}, achieving long-term stability better than \SI{1}{\micro\kelvin}.

With neutron-activated crystals, the LiFE-SNS detectors exhibited frequent $\beta$-decay events at a rate of approximately 40\,Bq. During the calibration campaign, the total event rate increased further due to additional $\gamma$-ray absorption events with larger amplitudes than most $\beta$ events. These high event rates modified the frequency spectrum of the randomly triggered signals, most noticeably in the low-frequency region, with different levels observed between calibration and science runs. These differences remained within the passband of [100, 1500]\,Hz, which had originally been selected based on the maximum signal-to-noise ratio. To mitigate this effect, an alternative bandpass filter of [900, 2000]\,Hz, featuring a higher high-pass cutoff, was applied to each signal. After applying this bandpass to noise signals acquired during both calibration and science runs, no significant differences were observed in either the filtered noise spectra or the averaged signal templates, thereby validating the calibration procedure.

In addition, noise conditions with enhanced low\hspace{0pt}-\hspace{0pt}frequency components can bias the reconstruction of low-energy events. This effect was evaluated by testing the amplitude\hspace{0pt}-\hspace{0pt}determination logic using pseudo-signals constructed by summing scaled templates of known amplitude with noise data sampled in random time windows. The resulting bias-correction method was found to be effective for low-energy events, particularly under conditions of high event rates. The adjusted amplitudes introduce a modified nonlinear behavior in the energy calibration and should be taken into account in the analysis.

Figure~\ref{fig:e_cal_h3} shows the calibration results measured with a detector setup employing the same crystal during the detector-characterization tests of Exp.~1--10. The $^3$H $\beta$-decay events dominate below 20 keV. Mn K$_{\alpha,\beta}$ X-ray lines from the internal source are clearly identified, along with the Ag K$_{\alpha,\beta}$ lines near their nominal energies. A W K$_\alpha$ peak is also visible above background when including a 13-day calibration run. Dotted vertical lines in the top panel of Fig.~\ref{fig:e_cal_h3} mark the energies of the identified peaks. Events corresponding to the Ag K$_\alpha$ line (22.16 keV) were used to generate a signal template for pulse-amplitude determination. The Mn K$_{\alpha,\beta}$ peaks are fitted with two independent Gaussian functions that share a common width parameter, together with a third-order polynomial describing the local background from $^3$H decays. The other three groups of peaks, Ag K$_\alpha$, Ag K$_\beta$, and W K$_\alpha$, are fitted with multiple Gaussian functions that share a common width parameter and have fixed relative
normalizations and mean values as provided by the NNDC database~\cite{NNDC}, together with an exponential background. The mean and standard deviation of the fitted Gaussian function for the most dominant peak in each group were used to estimate the energy calibration and resolution functions.

To correct for the position dependence of the peak amplitudes, additional Geant4 simulations were performed to estimate the event locations associated with each X-ray peak and to determine the corresponding $\overline{F}$ values. The solid black line in the middle panel of Fig.~\ref{fig:e_cal_h3} represents the energy calibration after applying the position correction described in the previous section. The amplitudes of the 6\,keV events from the internal source were shifted by a negligible amount with respect to the reference template at 22.16\,keV, unlike the higher-energy peaks. The observed nonlinearity is of a similar order, but with slightly different characteristics compared to the setup before neutron activation, originating from the increased event rate. This nonlinear behavior can be accounted for using a third-order polynomial calibration curve with a zero constant term, as shown by the solid curve.

Assuming that the total measured resolution $\sigma_\mathrm{measured}$ arises from independent contributions of the position\hspace{0pt}-\hspace{0pt}dependent effect ($\sigma_\mathrm{position}$), baseline noise ($\sigma_\mathrm{baseline}$), and other performance factors ($\sigma_\mathrm{others}$), it can be expressed as
\begin{equation}
\sigma_\mathrm{measured}^2 = \sigma_\mathrm{position}^2 + \sigma_\mathrm{baseline}^2 + \sigma_\mathrm{others}^2.
\label{eq:resol}
\end{equation}
Here, $\sigma_\mathrm{others}$, shown as empty magenta squares in Fig.~\ref{fig:e_cal_h3}, includes energy-dependent contributions arising from detector nonlinearity, temperature instability, gain drift, and related effects. 
It is obtained by subtracting the position-dependent contribution and the baseline resolution from the measured resolution in the calibration run, following Eq.~\ref{eq:resol}. A quadratic function without a constant term is fitted to $\sigma_\mathrm{others}$ and shown as the magenta dotted line in Fig.~\ref{fig:e_cal_h3}.

The position-related resolution, $\sigma_\mathrm{position}$, is calculated using the amplitude relation defined in Eq.~\ref{eq:amp}, and is represented by blue asterisks in Fig.~\ref{fig:e_cal_h3}. Specifically, $\sigma_\mathrm{position}$ on the amplitude scale is obtained from the standard deviation of the position-dependent response $F(x,y,z)$, as defined in Appendix~\ref{appendix}, multiplied by the position-corrected energy calibration function. To estimate $\sigma_\mathrm{position}$ for $^3$H events, the simulated $^3$H spatial distribution described in section ~\ref{subsec:tritiumGen} was used. The energy dependence of $\sigma_\mathrm{position}$ for $^3$H events is shown as a red dashed line in Fig.~\ref{fig:e_cal_h3}.

The baseline resolution, \(\sigma_\mathrm{baseline}\) (blue dash-dotted line in Fig.~\ref{fig:e_cal_h3}), is derived from the amplitude distribution obtained by injecting signal templates into noise data from the calibration campaign, following the same procedure as previously described in the bias correction method. No significant energy dependence is
observed.

Based on the above, the resulting $^3$H resolution function is shown in the bottom panel of Fig.~\ref{fig:e_cal_h3}. The black line
represents the final resolution, with the yellow shaded region indicating the 95\% confidence interval.
Notably, the quadratic sum of $\sigma_\mathrm{baseline}^2$ and $\sigma_\mathrm{others}^2$ dominates over $\sigma_\mathrm{position}^2$ for the $^3$H distribution in the LiF crystal, indicating that position\hspace{0pt}-\hspace{0pt}dependent effects contribute only marginally to the overall energy resolution.

%% file: sec5_systematics.tex
\section{Systematic Considerations}
\label{sec:systematics}

Using this detector technology for a sterile-neutrino search, we investigate possible mechanisms that may contribute to systematic uncertainties. The items discussed below should be accounted for in the expected $^3$H $\beta$ spectrum or considered as potential sources of systematic errors.

\subsection{Spectral shape of tritium beta spectrum}

In LiFE-SNS, $^3$H nuclei are embedded directly within a LiF crystal, realizing a source-equal-to-detector configuration for tritium $\beta$ decay. In this configuration, the detector measures the total decay energy, excluding the energy carried out by the neutrino. The measured spectrum therefore includes contributions from the $\beta$ electron, the recoil of the daughter nucleus, and atomic de-excitation and neutralization processes. 

These processes dissipate their energy in the crystal and contribute to the calorimetric signal. As a result, the measured spectrum reflects all the components, introducing potential systematic uncertainties for sterile-neutrino searches that rely on small spectral distortions around a keV-scale kink.

The dominant contribution arises from the $\beta$ electron. The relativistic three-body decay distribution for the electron and anti-neutrino can be expressed as~\cite{ludl2016jhep}
\begin{multline}
\frac{d{\mathrm{\Gamma}}}{dE_{e}}=
\frac{1}{64\pi^{3}m_{^3\mathrm{H}}}
\int_{E_{j-}}^{E_{j+}} dE_j\,
\left|M({^3\mathrm{H}} \to {^3\mathrm{He}} + e^{-} + \bar{\nu}_j)\right|^2,
\label{eq:nuSpect}
\end{multline}
where $E_j$ is the neutrino energy, $E_{j+}$/$E_{j-}$ are the upper/lower bounds of the neutrino energy for a given electron energy $E_e$ and mass of ${^3\mathrm{He}}$ including the binding energy of orbital electrons, $m_{^3\mathrm{H}}$ is the mass of tritium, and $\left|M\right|^2$, explicitly a function of $E_e$ and $E_j$, is the matrix element squared summed over particle spins. Theoretical studies have detailed the structure of $\left|M\right|^2$~\cite{simkovic2008prc,long2014jcap,mertens2015jcap,ludl2016jhep,leendert2018rmp}, accounting for the vector, axial-vector, and weak-magnetism contributions in the weak interaction. Equation~\eqref{eq:nuSpect} thus represents the differential decay rate as a probability distribution for neutrino energy given a fixed electron energy.

Despite extensive theoretical work on tritium $\beta$ decay, residual uncertainties remain in the relative intensities and detailed shapes of the additional atomic and solid-state corrections~\cite{mertens2015jcap,ludl2016jhep,leendert2018rmp,hayen2018rmp}. Moreover, atomic de-excitation and charge-neutralization processes occur faster than the thermal response of the LiFE-SNS detectors, and can be included in precision $\beta$-spectrum modeling. For a sterile-neutrino search, such effects manifest as potential biases in the spectral shape and kink amplitude, particularly in the full mass region relevant for LiFE-SNS.

\subsection{Surface-related energy loss} 

As shown in Fig.~\ref{fig:tritium-dist}, neutron simulations indicate a nonuniform distribution of $^3$H within the LiF crystal, with little directional dependence.
This tritium distribution is used in separate Monte Carlo simulations to estimate the energy deposited in the LiF crystal by electrons from $^3$H $\beta$ decay and to evaluate distortions of the spectral shape caused by nonmeasurable energy losses due to X-ray or electron escape near the crystal surface.

\begin{figure} [t]
\centering
\includegraphics[width=0.9\columnwidth]{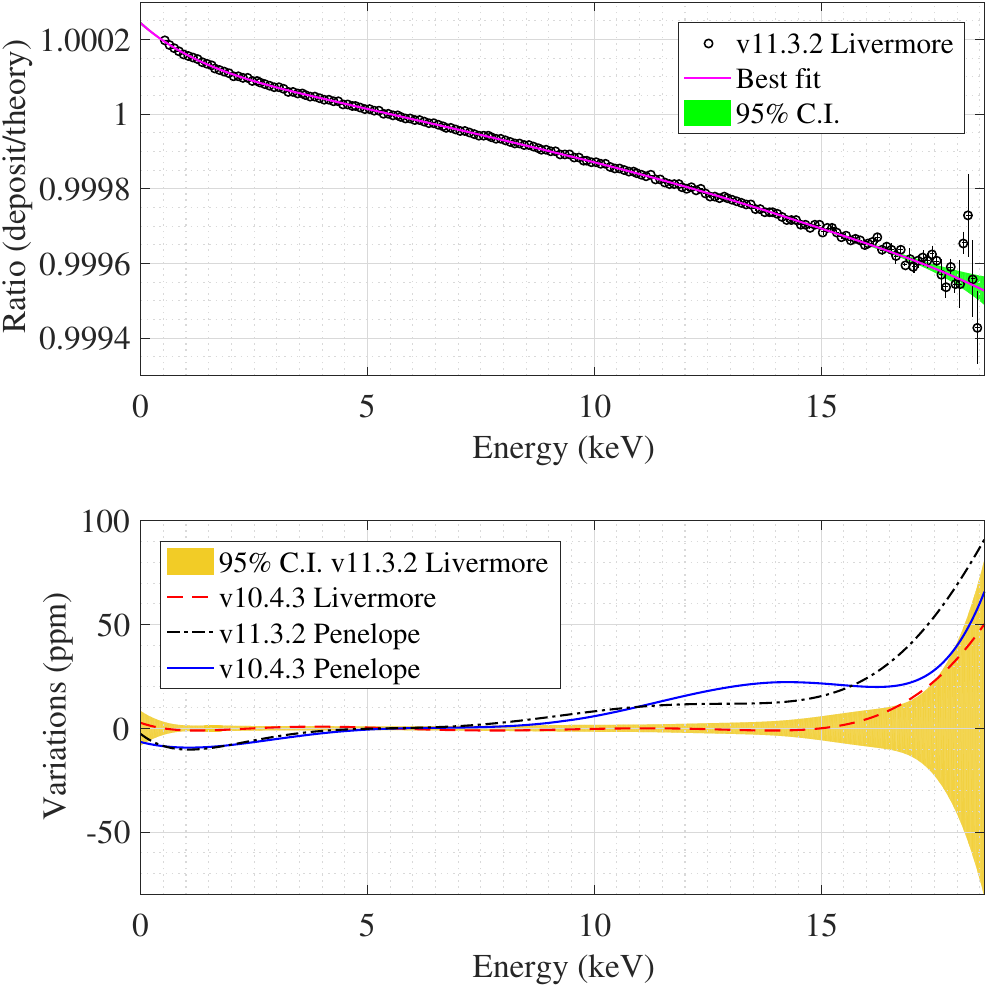}
\caption{(\textit{Top}) Ratio of the deposited energy spectrum obtained from the simulation to the assigned theoretical $^3$H spectrum. The reference result is derived from $5\times10^{9}$ generated $\beta$ events using Geant4-11.3.2 with the Livermore EM package. The solid line indicates a seventh-order polynomial fit to this reference result.
(\textit{Bottom}) Variations of the simulated spectra obtained with different Geant4 versions and EM physics packages relative to the reference line. The shaded band indicates the statistical 95\% confidence interval corresponding to $10^{9}$ generated $\beta$ events.}
\label{fig:surface-escape}
\end{figure}

Figure~\ref{fig:surface-escape} shows the distortion of the deposited-energy spectrum relative to the assigned $^3$H spectrum, obtained using several recent Geant4 versions and low-energy electromagnetic (EM) physics packages. The reference result, shown in the top panel, is derived from $5 \times 10^{9}$ generated $\beta$ events using Geant4-11.3.2 with the Livermore package. Additional simulations using Geant4-10.4.3 with the Penelope and Livermore packages employ five times fewer generated events. 

The ratio of the deposited spectrum to the theoretical spectrum exhibits a small but clear shift toward lower energies. The solid line represents a polynomial fit to the reference result, accompanied by narrow statistical confidence intervals. These results indicate that surface-escape effects can be reliably incorporated into the expected spectrum using a $^3$H depth profile modeled with recent Geant4 simulations.

The bottom panel shows variations of the other simulation results from the reference fit. The shaded region indicates the confidence interval corresponding to $10^{9}$ generated $\beta$ events. The three lines represent best-fit curves to the ratio data obtained using older Geant4 versions and different EM physics packages; their confidence intervals, comparable in size to that of the reference result, are omitted for simplicity. All results show no significant deviation from the reference simulation using the Livermore package in Geant4-11.3. Simulations using the same package with an older Geant4 version agree within statistical uncertainties, while results obtained with the Penelope package are mutually consistent but exhibit a small systematic offset relative to those using the Livermore package, corresponding to slightly reduced energy loss.

Since these differences are not significant at the current statistical level, variations among low-energy EM simulation packages are not expected to impact the data from the first LiFE-SNS phase. However, for future phases with substantially higher $\beta$ statistics, further validation of low-energy simulations or dedicated experimental measurements may be required.

\subsection{Unresolved pile-ups}

Two or more events may occur within a single analysis time window, producing pile-up signals. Most of these pile-up events can be identified and excluded during the analysis. However, a fraction remains unresolved and thus contributes to the measured spectrum. Although their rate and spectral shape can be estimated using pseudo-experiments and incorporated into the expected spectrum, the uncertainty associated with this estimation introduces a potential source of systematic error.

\subsection{Solid-state effects }

Because $^3$H nuclei produced after neutron capture occupy interstitial sites in the LiF lattice, solid-state effects may influence their $\beta$-decay spectrum. Interactions with neighboring Li and F ions can perturb the local electronic environment of $^3$H, potentially requiring further theoretical modeling for accurate spectral predictions~\cite{hayen2018rmp}

In addition, the crystal potential can modulate the decay probability, giving rise to oscillatory distortions in the energy spectrum. This phenomenon, known as beta environmental fine structure (BEFS)~\cite{koonin1991naure}, has been observed in the $\beta$ spectra of $^{187}$Re in superconducting rhenium and in the Re compound AgReO$_4$~\cite{gatti1999nature,arnaboldi2006prl,rehr2000theoretical}.

For $^3$H embedded in LiF, the BEFS amplitude is expected to be weaker than in rhenium-based systems because the tritium occupies interstitial rather than substitutional sites and interacts less coherently with the surroundings. The effect is expected to appear primarily in the low-energy region of the spectrum. As observed in Re-based crystals, BEFS modulations are most pronounced below 1\,keV, with oscillation periods less than 200\,eV. Observing such structure in LiF would likely require substantially better energy resolution than that achieved in the present work.

The BEFS effect has implications for searches for sterile neutrinos, as it may obscure or mimic a spectral kink around 1 keV, 
potentially misleading the analysis to infer a sterile-neutrino mass 1 keV less than the decay energy. 
A more detailed understanding of these solid-state effects will be pursued with a high-statistics measurement of the $^3$H $\beta$ spectrum in the low-energy region.

%% file: sec6_conclusion.tex
\section{Conclusion}

We have carried out a series of measurements to characterize the detector performance of the LiFE-SNS setup, which consists of a LiF crystal thermally coupled to an MMC sensor. The detector properties were systematically investigated for precision measurements of the $^3$H $\beta$-decay spectrum, demonstrating that this approach is well suited for searches for keV-scale sterile neutrinos.

Two detector modules, each incorporating a LiF crystal with embedded $^3$H, were prepared for the first stage of the LiFE-SNS experiment. A 13-day calibration campaign revealed characteristic X-ray lines, enabling the construction of reference signal templates and demonstrating reasonable energy resolution and well-characterized linearity over the observed $^3$H $\beta$-decay event rates.

Following the calibration campaign, the system operated stably during a four-month data-taking period. Assuming that systematic uncertainties remain substantially smaller than the statistical uncertainty for the full dataset, comprising two channels over four months, the expected sensitivity approaches the statistical limit shown in Fig.~\ref{fig:bounds}. This corresponds to nearly an order-of-magnitude improvement over existing constraints derived from $\beta$-decay spectrum measurements in the $\sim$10\,keV mass range. The analysis results from this dataset will be presented in future publications.

As discussed in this work, the preparation of $^3$H embedded LiF crystals is straightforward, and the detector performance is well understood. The achieved energy resolution is sufficient to probe subdominant features in the tritium $\beta$ spectrum. These results support the feasibility of scaling up the LiFE-SNS experiment, in particular by increasing the number of readout channels. The use of smaller LiF crystals within the same detection concept would further reduce signal rise and fall times, allowing higher activity per channel while mitigating unresolved pile-up effects. The projected sensitivity of an upgraded phase, assuming a three-year run with 100 channels operating at 100\,Bq per channel, is also shown in Fig.~\ref{fig:bounds}.

Once the KATRIN experiment transitions to wide-range energy measurements of the $^3$H $\beta$ spectrum with its TRISTAN detector upgrade~\cite{mertens2019jpg}, following the endpoint-search program, it is expected to achieve significantly higher statistics than those obtained in the present study or in the measured two-channel, four-month dataset. Nevertheless, the LiFE\hspace{0pt}-SNS project provides a complementary approach, employing the same tritium $\beta$-decay source but relying on a fundamentally different detector technology.

%% file: appendix.tex

\begin{appendices}

\section{Position calibration analysis}\label{appendix}

The present detector, consisting of a LiF crystal with a gold phonon-collector film deposited on one face, exhibits position-dependent variations in signal amplitude. These variations arise primarily from the transport of athermal phonons, which are initially generated at the interaction site and are collected more efficiently for events occurring closer to the gold film, in general~\cite{krwoo2024jltp}. Because the detector energy resolution is sufficient to resolve this position dependence in the measured energy spectrum, a position calibration is required for a precise measurement of the $^3$H $\beta$-decay spectrum over the full LiF crystal volume. This Appendix describes the procedure used to determine the position-calibration factor $F$ introduced in Eq.~\ref{eq:amp}.

\subsection{Reference event-position distributions}

To utilize the reference data obtained in Exp.~2--10, spatial distributions of X-ray absorption sites were generated using Geant4 (v10.4.3) with the Livermore package. As shown in Fig.~\ref{fig:6kev-pos}, the simulated 5.9\,keV reference events are distributed near the crystal surface for eight different source positions. Additional groups of spatial distributions corresponding to characteristic X-ray lines produced by the $^{241}$Am source and its Ag/Cu collimator were simulated in the same manner. For higher-energy interactions, such as the 59.54\,keV $\gamma$ rays, where multiple Compton scatterings are likely to occur, an energy-weighted mean absorption point was adopted to define the effective interaction position.

In total, 21 distinct event groups, with different energies and source geometries, were simulated and used as position-calibration references. For each group, more than $5\times10^{5}$ events were generated to determine the corresponding absorption\hspace{0pt}-\hspace{0pt}site distributions. 
Because the spatial density of simulated interaction points varies among the groups, the number of reference points retained for each group was adjusted to ensure balanced sampling in the subsequent interpolation procedure. 
This adjustment resulted in a composite reference set of 27{,}666 simulated points representing the reference events for experimentally observed peaks.

After applying the four-fold rotational and reflection symmetries of the detector geometry, each reference set comprised 221{,}328 points with known $(x,y,z)$ positions and associated amplitudes. To ensure statistical robustness, 10{,}000 independent reference sets were generated and used in the position-calibration study.

\subsection{Interpolation method}

To isolate the position-dependent term, the measured amplitudes for each of the 21 reference groups were normalized by the energy-dependent response, $\mathcal{E}(E) = aE^{2} + bE$, introduced in Eq.~\ref{eq:amp}. Each simulated point was therefore assigned a value of $F(x,y,z)_{\mathrm{ref}}$ obtained from the corresponding normalized amplitude. One complete reference set thus consisted of 221{,}328 values of $F(x,y,z)_{\mathrm{ref}}$ associated with known positions, which served as anchor points for interpolation.

A linear interpolation scheme was used to estimate $F(x,y,z)$ for arbitrary event positions. For a given point $(x,y,z)$, 10{,}000 interpolated values of $F_{\mathrm{intpl}}$ were generated, and the mean of the central 50\% of these values was adopted as the effective position-correction factor $F(x,y,z)$.  

This method was then applied to newly simulated event positions for each of the 21 reference groups. A trimmed mean of the resulting $F(x,y,z)$ values—retaining entries within two times the median absolute deviation—was calculated, producing a mean correction factor $\overline{F}$ for each group. These $\overline{F}$ values were subsequently used to refine the calibration parameters $a$ and $b$ for all the 21 groups.

The procedure was iterated until the energy calibration parameters obtained before and after applying the position correction agreed to within 99.99\%. Convergence was typically achieved within a few iterations and determines the final form of the energy-dependent response $\mathcal{E}(E)$. For any event-distribution group, the corresponding mean correction factor $\overline{F}$ can then be computed using the same interpolation and trimmed-mean method.

\begin{figure} [t]
\centering
\includegraphics[width=1.0\columnwidth]{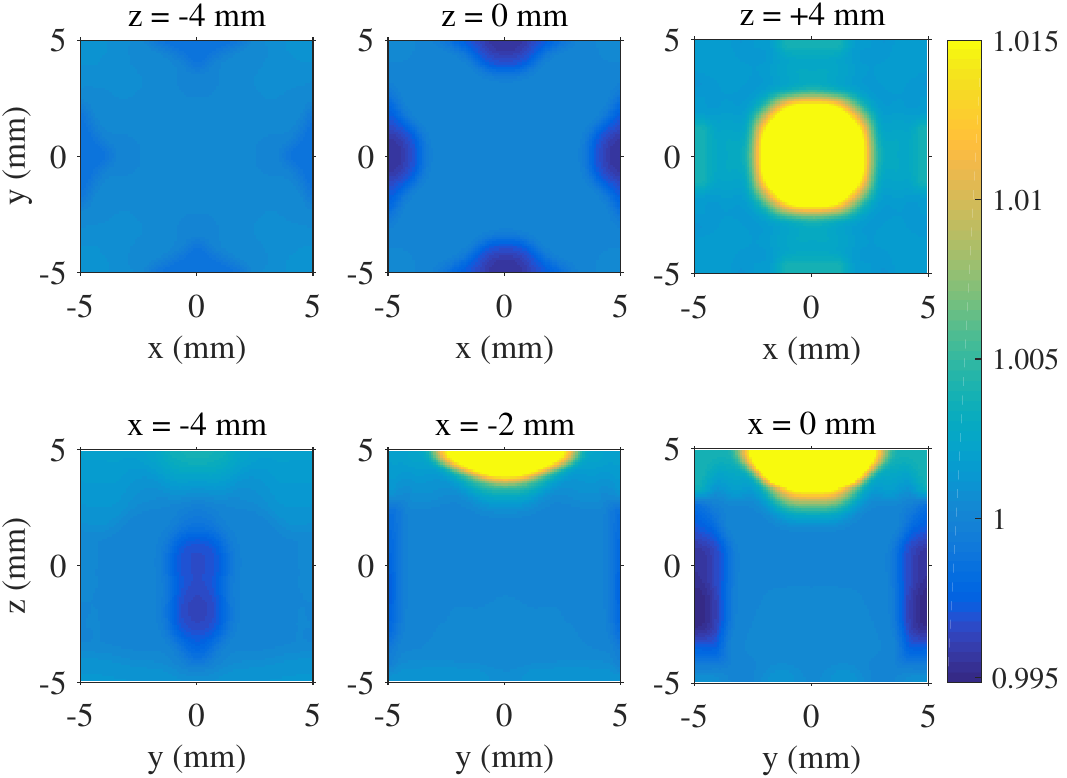}
\caption{Cross-sectional views of the position-calibration results for $(x,y,z)$ distributions localized within $(0.1\,\mathrm{mm})^3$ voxels across the cubic crystal volume.}
\label{fig:pos-result-pcolor}
\end{figure}

\subsection{Result}

Figure~\ref{fig:pos-result-pcolor} presents validation results for the position calibration method, obtained by evaluating $\overline{F}$ for $(x,y,z)$ distributions localized within $(0.1\,\mathrm{mm})^3$ voxels across the 1\,cm$^{3}$ crystal volume.
Distinct amplitude features corresponding to different $^{55}$Fe source locations in  Exp.~2--10 are clearly visible in the cross-sectional maps of $\overline{F}$.

As discussed earlier in Section \ref{subsec:position-calib}, the results of applying this position calibration to all reference peaks are shown in the right panel of Fig.~\ref{fig:pe-cal}. Most data points closely follow the fitted quadratic calibration curve, indicating that the dominant position-dependent effects have been effectively mitigated. The resulting mean correction factor $\overline{F}$ thus provides a practical position calibration that can be applied on an event-averaged basis when the
event-position distribution is available.

\end{appendices}